# Incentive Decision Processes

**Sashank J. Reddi**
Machine Learning Department
Carnegie Mellon University
sjakkamr@cs.cmu.edu

**Emma Brunskill**
Computer Science Department
Carnegie Mellon University
ebrun@cs.cmu.edu

## Abstract

We consider Incentive Decision Processes, where a principal seeks to reduce its costs due to another agent's behavior, by offering incentives to the agent for alternate behavior. We focus on the case where a principal interacts with a greedy agent whose preferences are hidden and static. Though IDPs can be directly modeled as partially observable Markov decision processes (POMDP), we show that it is possible to directly reduce or approximate the IDP as a polynomially-sized MDP: when this representation is approximate, we prove the resulting policy is boundedly-optimal for the original IDP. Our empirical simulations demonstrate the performance benefit of our algorithms over simpler approaches, and also demonstrate that our approximate representation results in a significantly faster algorithm whose performance is extremely close to the optimal policy for the original IDP.

## 1 Introduction

Consider a landlord who pays for her tenant's heating bill, or an insurance company that covers an insuree's health care bills. These are two instances where a principal agent incurs a cost for the behavior of another agent. In such scenarios the principal may be able to provide incentives to attempt to change the agent's behavior; however, the principal typically does not know the preferences of the agent, and these preferences determine which incentives will be accepted by the agent in return for altered behavior.

We formalize this problem as an instance of sequential decision making under uncertainty. Despite much AI interest in this general field, there has been limited research on multi-step decision processes where a principal agent's own costs are a function of another agent's behavior. We call these processes Incentive Decision Processes (IDPs) and in this paper we focus on the case where a principal interacts with a greedy myopic agent whose preferences are hidden and static. The objective is to compute a decision policy for the principal that minimizes its total expected sum of costs over the horizon of interactions with the agent. Computing this policy would be trivial if the agent's preferences were known to the principal: the principal would simply offer the incentive for an agent action that minimized the principal's cost amongst all incentive-action pairs that would be accepted by the agent. However, the agent preferences are typically hidden. It is important to consider the case where a principal cannot simply ask the agent for its preferences. Naturally a strategic rational agent may wish to misrepresent its stated preferences in order to gain additional reward, and we will briefly touch on this issue at the end of the paper. But our focus will be on agents that act myopically and therefore truthfully (as we will later prove). We argue that this model is reasonable for many important situations where the agent is not adversarial, but is either unaware or in denial of its own utility function, such as asking a person about their preference for consuming healthy food or exercising different amounts, or where it is impossible for the agent to directly respond (such as teaching a baby a new behavior).

Our work is most closely related to the research of Zhang and colleagues [12, 13, 4] on environment design. However, Zhang et al.'s work primarily addresses inducing the agent to follow a particular policy within a short number of interactions, whereas our research focuses on how to provide incentives to minimize the total cost incurred by the principal over many interactions with the agent. We believe our objective is likely to be particularly relevant in multiple real-world domains, where the cost of any diagnostic steps to reveal the hidden preferences of the agent, such as a tenant, office employee, or insuree, must be balanced with the expected benefit of that information in reducing the cost to the principal. In this sense our algorithms will address the exploration versus exploitation tradeoff in the context of Incentive Decision Processes.

We will show constructing a policy for the principal in an IDP may be modeled as a partially observable Markov decision process (POMDP) planning problem. However we will demonstrate that we can leverage the significant structure in IDPs to achieve much more efficient optimal or boundedly-optimal algorithms. Our first contribution is to prove that if there are only two different actions that the agent may choose, then the incentive decision process can be reduced to a polynomially sized MDP. We will also describe a similar approximate mapping for multiple action incentivized decision processes, and also present formal bounds on the quality of policies for the approximate representation. Our empirical simulations demonstrate the performance benefit of our sequential decision-theoretic algorithms over simpler approaches. We will finish by briefly discussing multiple extensions to our model, including dynamic agent preferences and strategically rational agents. When proofs are omitted, they are provided in the appendix.

## 2 Background

A Markov decision process (MDP) is described by a tuple $\langle S, A, p(s'|s,a), C, \gamma \rangle$. $S$ is a set of states, $A$ is a set of actions, and $p(s'|s,a)$ is the transition model and stores for each state $s$ and action $a$ the probability of transitioning to any state $s'$. $C$ is the cost model and represents the cost of taking action $a$ in state $s$ and transitioning to state $s'$. $\gamma \in [0,1]$ is a discount factor. A stationary policy $\pi : S \to A$ is a mapping from states to actions.[1] The value of a policy $\pi$ for a horizon $H$ from state $s$ is the expected sum of costs over $H$ steps when following the policy from state $s$,

$$V_\pi^H(s) = \sum_{s'} p(s'|s,\pi(s))[c(s,\pi(s),s') + \gamma V_\pi^{H-1}(s')]$$

This equation is also known as the Bellman equation [1]. In a slightly modified form (known as a Bellman update) it can be used to iteratively compute the optimal value function and optimal policy. If the horizon is infinite, then the value function is independent of the time step. The optimal policy is the policy that has the lowest value.

Partially observable MDPs (POMDPs) [10, 6] are often used to describe decision processes where the state is not directly observable. POMDPs are represented by a tuple $\langle S, A, O, p(s'|s,a), p(o|s',a), C, \gamma \rangle$. Here $O$ is a set of observations, and $p(o|s',a)$ is the probability of observing $o$ after taking action $a$ and transitioning to state $s'$. A belief $b$ is a probability distribution over the possible world states, given the prior history of actions taken and observations received. $b$ is a sufficient statistic for the history, and can be updated as new actions are taken, and new observations are received, using a Bayes filter. In POMDP planning the objective is to compute a policy $\pi : b \to a$ that maximizes the expected sum of future rewards.

## 3 Related Work

Since the agent's preferences are unknown to the principal, our work is loosely related to research on preference elicitation [3]. However, preference elicitation only focuses on identifying an agent's hidden reward model, or identifying it sufficiently to make a single decision. In contrast, in our work the principal seeks to minimize its expected sum of costs. Similarly, inverse reinforcement learning [9] seeks to identify an agent's hidden reward model given the agent's behavior, but our focus is on minimizing the cost to a principal by altering the agent's behavior through the offering of incentives, which may not involve fully identifying the agent's hidden reward model.

Research on multiple agents interacting includes sophisticated models for representing and updating each agent's internal estimate of the state of the other agent. Such representations can lead to infinite nestings of state estimates [11]. Interactive POMDPs (I-POMDPs) are used to represent such problems, and use a finite representation of each agent's state for tractability [5]. Such rich representations remain very expressive, but at the price of computational intractability for all but very small domains.

The closest work to our own is that of Zhang and colleagues [12, 14, 13, 4] on policy teaching and environment design. Their earliest work [12] focuses on finding a set of incentives for the agent that minimizes the principal's expected cost. Their setting is quite general: the agent acts in a MDP and the principal provides a set of incentives over the MDP states. Zhang et al. provide an algorithm for identifying the optimal incentives to provide within a finite (but unknown) number of episodes, where in each multi-step episode the agent follows their optimal policy given their (hidden) reward function plus the provided incentives. Our presented work focuses on a more restrictive setting of a greedy agent selecting among a fixed set of actions at each step, but within this context we provide formal guarantees of our algorithm's performance over a *single* multi-step episode as the principal and agent interact. Our IDP model is most similar to Zhang et al.'s recent work [4] on an incentivized multi-armed bandit. However, this work focuses on inducing a particular action selection by the agent given some budget, whereas our focus is on minimizing the total expected cost for the principal. Also, we are interested in long horizon decision processes, versus their algorithmic work only considers short ($\leq 3$ step) horizons.

---
[1] The policy can also depend on the time step.

## 4 Incentive Decision Process Model

Consider a principal interacting multiple times with an agent. At each time step the agent chooses an action from a set of agent actions $A = \{a_1, a_2, \cdots, a_N, a_{N+1}\}$. We assume that the agent has an internal reward $R_{a_n}$ for every action $a_n$ such that $R_{a_i} > R_{a_j}$ if $i > j$. Therefore $a_{N+1}$ will yield the highest reward for the agent, and so we define $a_{N+1}$ as the default agent action that the agent will select if no incentives are provided. All other actions $a_1, \cdots, a_N$ will be referred to as alternate agent actions. The agent's rewards are hidden from the principal, but due to the assumed structure, the principal does know that action $a_{N+1}$ is the agent's default action. Each agent action $a_i$ is also associated with a particular cost $c_i$ to the principal and the costs are such that $c_i < c_j$ if $i < j$. Therefore by definition the default agent action $a_{N+1}$ is of highest cost to the principal.

At each time step the principal offers an incentive $\delta$ for a particular action $a_n$ ($n \in (1, N)$) to the agent if the agents chooses action $a_n$ on this time step. The offered incentive is selected from a set of $K$ incentives $\Delta = \{\delta_1, \cdots, \delta_K\}$ where $\delta_i < \delta_j$ if $i < j$. We assume that the agent acts to maximize its immediate reward, and will accept the offered incentive and take action $a_n$ if doing so yields a higher reward than the default action, $R_{a_n} + \delta > R_{a_{N+1}}$. Otherwise the agent will reject the incentive and take the default action $a_{N+1}$ and receive reward $R_{a_{N+1}}$. If the agent accepts the offered incentive for action $a_n$, the principal will incur an immediate cost of $c_n + \delta$; otherwise, the principal's immediate cost is $c_{N+1}$. We will assume that the cost of each alternate agent actions, plus the maximum possible incentive, is always less than the cost of the default agent action: $c_N + \delta_K < c_{N+1}$. This implies that the principal would always prefer the agent to accept an incentive and take one of the alternate agent actions. We also assume that for each alternate agent action $a_n$, there exists an incentive in $\Delta$ such that if that incentive is offered, the agent would prefer to accept that incentive and take the alternate agent action instead of the default action. Let $t_n = R_{a_{N+1}} - R_{a_n} \in \Delta$ be the least incentive for which the agent prefers action $a_n$ to the default agent action $a_{N+1}$. Observe that lower indexed actions have lower rewards to the agent, and therefore will require incentives equal or greater than higher indexed rewards, $t_i \geq t_j$ for $i < j$. We use $I = (t_1, \cdots, t_N)$ to denote the tuple of true incentives for the $N$ alternate agent actions. These incentives are initially unknown to the principal. The principal is provided with an initial joint probability distribution $P_0$ which gives the probability that $t_i = \delta_\mathbf{k}$ for all $n \in \{1, \cdots, N\}$ and $k \in \{1, \cdots, K\}$.

Note the set of alternate actions and set of possible internal rewards for each action for the agent is common knowledge. The agent's actual reward for each alternate action is prive knowledge to the agent. Since the agent is assumed to act in a myopic fashion, it does not matter if the principal's costs for each alternate action, and its set of possible incentives, are private or common knowledge.

The objective of the principal is to minimize its expected sum of costs over $H$ interactions with the agent. The costs may be multiplied by a discount factor $\gamma \in [0, 1]$. We will consider both the finite horizon case and the discounted infinite horizon case ($H = \infty$). In this paper we will design tractable algorithms that compute a decision policy to achieve the principal's objective.

For example, the landlord-tenant scenario can be modeled as an IDP where the tenant's (agent's) default temperature setting is the default action, and other temperatures correspond to alternate agent actions. Each temperature maps to a cost to the landlord (principal). At each step the principal offers an incentive to the agent if the agent sets the thermostat to an alternate temperature that the principal specifies.

An IDP can also be modeled as a POMDP. In the IDP POMDP the state is the agent's hidden $t_n$ for all $n \in \{1, \cdots, N\}$, the actions are the cross product of all alternate agent actions with the possible incentives $\Delta$ (since the principal can offer any incentive paired with any alternate agent action), and the observations are the binary accept or reject response of the agent. For most of this paper we assume that the agent's hidden preferences are static, and so the state transition model is a delta function. The observation model is that an agent's acceptance of an offer of $\delta_k$ for alternate action $a_n$ indicates that $t_n \leq \delta_k$ (since an agent will accept any incentive equal or greater than $t_n$ for alternate action $a_n$), and a reject indicates that $t_n > \delta_k$. Note that the observation model depends on how we assume the agent acts, which is a function of its internal hidden reward and the offered incentive. Since we assumed that the agent is myopically greedy, the agent will accept an incentive if the sum of offered incentive and the agent's internal reward is greater than the agent's hidden reward for its default action choice. The initial belief in the POMDP is given by $P_0$. Computing the optimal policy for this POMDP will enable the principal to optimally offer $(a_n, \delta_k)$ pairs that balance reducing the uncertainty over the agent's hidden true incentives $I$ with minimizing the expected cost to the principal, given the current belief state over $I$.

However, computing an optimal policy for a POMDP can take doubly exponential time and existing methods often struggle to scale to large state or action spaces or long time horizons, all of which can feature in IDPs.

Fortunately there is significant structure in IDPs. We will now show how we can leverage this structure to reduce the problem to simpler decision processes, and we will present efficient algorithms for computing exact or boundedly-optimal policies.

## 5 Single Alternate Agent Action IDPs

We first consider IDPs when there is a single alternate action $a_1$. It is known that a POMDP can be reduced to a continuous state MDP where each state represents a belief: note that since the states are probability distributions, the MDP has an infinite state space. Here we show the surprising result that a single alternate agent action IDP can be reduced to a polynomially sized MDP, which means they can be solved much more efficiently than standard POMDPs.

We first describe how to perform belief updating in a single alternate agent action IDP. Let the possible range of incentives for alternate agent action $a_1$ be $S_1 = [s_1, e_1]$ if and only if $\delta_{s_1} \leq t_1 \leq \delta_{e_1}$. $S = [S_1, \cdots, S_N]$ are the incentive ranges for all alternate agent actions; for $N = 1$, $S = S_1$. We are interested in maintaining the probability that $t_1 = \delta_i$ for all $\delta_i \in \Delta$. Given the initial probability distribution $P_0$, we can directly compute the initial incentive range $S_1$. The initial probability that $t_1 = \delta_i$ is

$$p_{1,S}(\delta_i) = \frac{P_0(\delta_i)\mathbb{1}(\delta_i, S)}{\sum_{k=1}^{K} P_0(\delta_k)\mathbb{1}(\delta_k, S)}, \quad (1)$$

where $\mathbb{1}(\delta_i, S) = 1$ if $\delta_{s_1} \leq \delta_i \leq \delta_{e_1}$ (if incentive $\delta_i$ is within the possible incentive range), and, with a slight abuse of notation, we use $P_0(\delta_i)$ to represent the initial probability that the agent's $t_1 = \delta_i$. Note that initially the denominator will be 1, because the range will cover all incentives $\delta_i$ for which $P_0(\delta_i) > 0$. On the first time step the principal will offer the agent an incentive $\delta_i$ to take the alternate agent action $a_1$. If the agent accepts, the principal knows that $t_1 \leq \delta_i$, due to the assumed monotonic structure of the incentives and the greedy myopic agent behavior. Therefore the new possible incentive range if the agent accepts $\delta_i$ is $S_1 = [s_1, i]$, since the upper bound on the possible index is at most $i$. If the agent rejects the offered incentive, then $t_1 > \delta_i$ and so the new incentive range must be $S_1 = [i+1, e_1]$. In either case the new probability of the agent accepting a particular incentive $\delta_j$ is identical to Equation 1 using the updated incentive range $S_1$.

Each offered incentive, and associated agent response, can set the probability of some incentives to zero. Since $t_1$ is static, we prove it is sufficient to maintain and update $S_1$ in order to compute the current $p_{1,S}(\delta_j)$ given the history of offered incentives and agent responses. Therefore, the belief state over the agent's incentive $t_1$ will be of the form $B_{i,j}^1 = (0, \cdots, 0, p_{i,S}^1, \cdots, p_{j,S}^1, 0, \cdots, 0)$ where $S_1 = (i, j)$ and $p_{j,S}^1 = p_{1,S}(\delta_j)$. We will shortly prove that there are only $O(K^2)$ possible such states.

Before we do so, we first note that an IDP-MDP is an instance of a deterministic POMDP, since the observations are a deterministic function of the underlying agent's preferences, and the transition model is a delta function because these preferences are assumed to be static. Prior work [7, 2] has proved that a deterministic POMDP with $N_s$ states can be converted to a finite-state MDP with a number of states that is an *exponential function* $((1 + N_s)^{N_s})$ of the number of POMDP states. This is significantly smaller MDP than generic POMDPs which map to a MDP with an infinite number of states. However, we will show that an IDP with $N = 1$ can be mapped to a MDP with $B_{i,j}^1$ as the MDP states. We now prove that the resulting MDP has a number of states that is only a *polynomial* function of the number of POMDP states $(K)$.

**Theorem 5.1.** *An IDP with $N = 1$ alternate agent actions is equivalent to an MDP with $O(K^2)$ states.*

*Proof.* We see that $B_{1,K}^1$ is the initial state of the IDP-MDP. Suppose $B_{i,j}^1$ is the current state. Consider the case where incentive $\delta_k$ is offered. The agent might either accept or reject the incentive. As stated, the agent's response to each offer can eliminate certain range of the incentives but the ratio amongst the remaining ones does not change. If the agent accepts the incentive, the new probability vector will be $B_{i,k}^1$ since the response indicates that $t_1 \in [\delta_i, \delta_k]$, otherwise if the agent rejects the incentive, the new probability vector will be $B_{k+1,j}^1$, since $t_1 \in [\delta_{k+1}, \delta_j]$. Hence, by induction, at the $t^{th}$ step, the probability vector is of the form $B_{i,j}^1$ and only states of this form are reachable from the initial probability vector. Now $i$ and $j$ can each take on at most $K$ values, so the maximum number of possible $B_{i,j}^1$ is at most $O(K^2)$. Therefore the number of belief states is bounded by $O(K^2)$ and therefore the IDP with one alternate agent action can be reduced to a MDP with $O(K^2)$ states. □

We now define the IDP-MDP for a single alternate agent action. We use $v(l)$ to denote the $l$-th element of a vector $v$. The state space is the $B_{i,j}^1$ vectors just described. The action space is the incentive space $\Delta$. The MDP transition model is the probability of transitioning from one $B_{i,j}^1$ to other belief states after the principal offers the agent an incentive. For a given state $B_{i,j}^1$, and a given offered incentive $\delta_k$, with probability $\sum_{i \leq l \leq k} B_{i,j}^1(l)$ the agent will accept the offer and the new state will be $B_{i,k}^1$, and with probability $1 - \sum_{i \leq l \leq k} B_{i,j}^1(l)$ the agent will reject the offer, and the new state will be $B_{k+1,j}^1$. The cost model of the MDP takes the form $c(B_{i,j}^1, \delta_k, B_{i,k}^1) = \delta_k + c_1$ for transitions where the agent accepts the incentive, and $c(B_{i,j}^1, \delta_k, B_{k+1,j}^1) = c_2$ when the agent rejects.

**Algorithm 1** ValH1
1: **Input:** IDP-MDP $M$, state $B_{i,j}^1$, matrix $V$ of horizon-values, matrix $\pi$ of horizon-policy, $h$
2: **if** $h = 0$ **then**
3: $\quad V^h(B_{i,j}^1) = 0$ {no more time steps}
4: **else**
5: $\quad$ **for** $k = i : j - 1$ **do**
6: $\quad\quad$ **if** $V^{h-1}(B_{i,k}^1) = \emptyset$ **then**
7: $\quad\quad\quad [V^{h-1}(B_{i,k}^1), V, \pi]$=ValH1($M, B_{i,k}^1, V, \pi, h$–1) {Calc. value if agent accepts $\delta_k$}
8: $\quad\quad$ **end if**
9: $\quad\quad$ **if** $V^{h-1}(B_{k+1,j}^1) = \emptyset$ **then**
10: $\quad\quad\quad [V^{h-1}(B_{k+1,j}^1), V, \pi]$=ValH1($M, B_{k+1,j}^1, V, \pi, h$–1) {Calc. value if agent rejects $\delta_k$}
11: $\quad\quad$ **end if**
12: $\quad\quad Q^h(B_{i,j}^1, \delta_k) = p(B_{i,k}^1 | B_{i,j}^1, \delta_k)(\delta_k + V^{h-1}(B_{i,k}^1)) + p(B_{k+1,j}^1 | B_{i,j}^1, \delta_k)(c_2 + V^{h-1}(B_{k+1,j}^1))$
13: $\quad$ **end for**
14: $\quad Q^h(B_{i,j}^1, \delta_j) = \delta_j h$
15: $\quad V^h(B_{i,j}^1) = \min Q^h(B_{i,j}^1, \cdot)$
16: $\quad \pi^h(B_{ij}^1) = \arg\min Q^h(B_{i,j}^1, \cdot)$
17: **end if**
18: Return $[V^h(B_{i,j}^1), V, \pi]$

We now show the special structure of this problem allows us to introduce more efficient planning algorithms that require only $O(K^3)$ and $O(min(H,K) K^3)$ computation time for the infinite and finite horizon case, respectively. First note that the principal only needs to consider offering incentives $\delta_k$ within the range of $i \leq k \leq j$ for a state $B_{i,j}^1$, as other incentives will be automatically rejected (or accepted but at higher cost than necessary for the principal). We start by considering the infinite horizon case. Let the expected sum of discounted costs of state $B_{i,j}^1$ be $V(B_{i,j}^1)$. We define a MDP state $s$ to be non-recurring with respect to a policy $\pi$ if the state can only be reached once while executing the $\pi$. A state $s$ of the MDP is called self-absorbing if executing $\pi$ from $s$ results in a self-transition to state $s$ with probability 1. A policy $\pi$ is NONRECUR-ABSORB if each MDP state is either non-recurring or self-absorbing with respect to $\pi$.

First recall that for an infinite horizon MDP, there always exists an optimal policy that is stationary (the decision depends on only the current state and is independent of the time step). Therefore we will consider only stationary policies. Next note that for an infinite horizon IDP-MDP, if incentive $\delta_j$ is offered in state $B_{i,j}^1$, then the MDP remains in same state $B_{i,j}^1$. If one of the other possible actions (offering incentives $\delta_i, \cdots, \delta_{j-1}$) is taken, then the MDP will transition to a new state, and will never return to the state $B_{i,j}^1$ since the number of non-zero entries in the probability vector $B_{i,j}^1$ monotonically decreases. This implies all infinite-horizon policies for an IDP-MDP are NONRECUR-ABSORB. As each state can only be reached once (or is an absorbing state) during execution, we only need to compute a Bellman backup for the state-action values $Q$ once for each state-action pair. We compute a decision policy by computing the state-action values of the initial state by recursively computing the value of each reachable state. The values of subsequent states are stored and cached so that they can be re-used if the same state is reached through a different trajectory of offered incentives and responses. There are $O(K^2)$ MDP states and $O(K)$ actions for each state, so there are $O(K^3)$ state-action pairs. The cost of computing the Bellman backup for a given state-action pair can be done in constant time given the values of the possible next states (since there are only at most 2 possible next states). Therefore the algorithm takes $O(K^3)$ time. Note this is significantly faster than value iteration on generic MDPs with $N_s$ states and $N_a$ actions which requires $O(\frac{N_a N_s^2}{1-\gamma})$ time to compute a near-optimal policy.

For the finite horizon case, let $V^h(B_{i,j}^1)$ denote the expected sum of costs to the principal for state $B_{i,j}^1$ for a horizon of $h$ future interactions with the agent. Every policy is not NONRECUR-ABSORB because the principal could offer the same incentive for multiple time steps (staying in the same state) and then offer a new incentive and transition to another state. However, we can prove the following result:

**Theorem 5.2.** *There exists an optimal policy for an IDP-MDP with $N = 1$ that is* NONRECUR-ABSORB.

The proof intuition is that we can always re-order an optimal policy so that it only offers incentives that keep the MDP in the same state at the final time steps.

The algorithm for the finite horizon case is displayed in Algorithm 1. Here the state's value and policy will depend on the time step. Each time the horizon decreases, either a state becomes self-absorbing with a cost of $\delta_j h$ or least one non-zero element of the probability vector $B_{i,j}^1$ becomes zero. Since the initial state $B_{1,K}^1$ has $K$ non-zero elements, it will take at most $K$ time steps until a state is reached which is self-absorbing (as all single-element beliefs are self-absorbing). Therefore, the algorithm takes $O(min(H, K) K^3)$ running time

## 6 Multiple Alternate Action IDPs

We next consider IDPs with multiple alternate agent actions, namely $N > 1$. We now need to maintain a belief state over the joint of the alternate agent actions crossed with the incentives. As stated in the definition of the IDP, we assume we are provided with an initial probability distribution over the true incentives $t_n$ for each alternate action $a_n$, $P_0$. We know from the prior section that all beliefs for an $N = 1$ IDP are of the form $B^1_{i,j} = (0, \cdots, 0, p^1_{i,S}, \cdots, p^1_{j,S}, 0, \cdots, 0)$ where $S_1 = (i, j)$. We now extend the model to handle multiple actions. Let $B_S = (B^n_S)|_{n=1}^N$ where $B^n_S$ denotes the probability vector $(0, \cdots, 0, p^n_{i,S}, \cdots, p^n_{j,S}, 0, \cdots, 0)$ where $S_n = (i, j)$ and $p^n_{i,S}$ is the probability that $t_n = \delta_i$ given the initial joint distribution $P_0$ and the possible incentive ranges $S_n$ for each alternate agent action. $S_n = (s_n, e_n)$, for all $n \in \{1, \cdots, N\}$.

**Theorem 6.1.** *The number of states of* IDP-MDP *for* $N > 1$ *is* $O(K^{2N})$ *states.*

*Proof.* Assume that the current state is $B_S$, where $S_n = (s_n, e_n)$ for $1 \leq n \leq N$. Consider the case where incentive $\delta_i$ is offered for action $a_n$. The agent will accept this offer with the probability that its true incentive $t_n$ for $a_n$ is $\leq \delta_i$, which is equal to $\sum_{j \leq i} p^n_{j,S}$. If the offer is accepted, then $s_n \leq t_n \leq i$, which implies $S_n = (s_n, i)$. In addition, all alternate agent actions $m > n$ with higher agent reward than alternate agent action $a_n$ will also all accept $t_n$, so the incentive range of all such actions will also be updated $S_m = (s_m, i)$ for all $m > n$. This defines a new $S$, which together with $P_0$, completely defines the new distribution over the probability of each incentive for each alternate agent action. If the agent rejects the offered incentive, then $s_n \geq i + 1$, and so $S_n = (i + 1, e_n)$. All alternate agent actions $m < n$ with a lower agent reward than action $a_n$ will also reject the reward, which results in $S_m = (i + 1, e_m)$ for all $m < n$. There are $N$ alternate agent actions, and we know from Theorem 5.1 that there are at $O(K^2)$ states per individual action, therefore there are at most $O(K^{2N})$ possible states. Due to the relationship among the agent costs and rewards, this number will often be much lower. □

An IDP with $N > 1$ alternate agent actions has many of the same properties as an IDP with $N = 1$ actions, and therefore we can use algorithms similar to those described in the prior section (see the text and Algorithm 1) to solve the IDP-MDP with $N > 1$. Excluding the cost for the belief updates, this requires $O(NK^{2N+1})$ and $O(min(H, NK)K^{2N+1})$ computation time for the infinite and finite horizon case, respectively. While this is computationally tractable for small $N$, this scales poorly as $N$ increases.

To address this, we will now provide efficient approximate algorithms for computing a decision policy in IDPs with large $N$. Note that computing the probability of the next possible states, given an offered incentive $\delta_k$ for alternate agent action $a_n$ involves computing the marginal probabilities $p^n_{k,S}$ that $t_n = \delta_k$ given the current $S$. This is an $O(K^N)$ operation due to summing over all other action-incentive probabilities. Hence, any approximation algorithm will take at least $O(K^N)$ time due to the bottleneck of belief updating. This cost can be reduced by assuming structure in the joint probability distribution, such as using graphical models with bounded treewidth. We leave further exploration of this issue for future work.

We now define a new special form of a Markov decision process. Let a SEQ-MDP be an IDP-MDP with $N > 1$ with the following additional restrictions on the allowable principal's actions at different states:

1. For the initial state, the principal can offer any incentive for only alternate agent action $a_1$ (assuming that initially $S_1 = (s_1, e_1)$ and $s_1 \neq e_1$).

2. For any other state, the principal can offer any incentive for alternate agent action $a_n$, only when $S_i = (s_i, e_i)$ and $e_i = s_i$ for all $i < n$. In other words, $t_i$ must be known for all alternate agent actions $i < n$ before the principal can offer an incentive for $a_n$.

A SEQ-MDP restricts the possible decision policy set $\Pi$ to policies that only provide incentives for alternate agent actions with higher cost to the principal only after finding the true incentive for the lower cost actions. Note that the SEQ-MDP policy may not identify the true incentive for all actions, but can decide to stick with a previously identified $a_n, \delta_k$ pair.

We now analyze the number of states in a SEQ-MDP. Consider that at the present state the principal is offering incentives for alternate agent action $a_n$. We know that $S_i = (s_i, e_i)$ for $i > n$ and $s_i = e_i$ for $i < n$. Define $c_i + \delta_j$ as the minimum action cost + true incentive across all $i < n$. Since we would never take any other $i' < n$ with a higher cost, $c_i + \delta_j$ is sufficient to summarize all useful information from prior offered actions. Therefore we can represent a state in the SEQ-MDP as the tuple $(a_i, \delta_j, B_{S_n})$. For each offered alternate agent action, there can be $O(K^2)$ states, as in the IDP-MDP with $N = 1$. We also must track the current minimal cost alternate agent action and offered incentives, and there are $O(NK)$ such combinations. Finally we also have to monitor the current alternate agent action, and there are $O(N)$ possibilities. The product of these quantities yields a state space of $O(N^2K^3)$. For similar arguments as given in Section 5, we can construct optimal policies that are NONRECUR-ABSORB for SEQ-MDP and use planning methods similar to those described in Section 5. There are $O(K)$ possible actions for each state in a SEQ-MDP. Therefore, again excluding costs for computing the marginal probabilities, the computational cost of computing an optimal infinite-horizon policy for a SEQ-MDP is $O(N^2K^4)$, and the finite horizon $H$ cost is $O(\min(H, NK)N^2K^4)$.

We now prove that computing the optimal policy for a SEQ-MDP yields a boundedly-optimal policy for the original IDP-MDP with $N > 1$.

**Theorem 6.2.** *An optimal policy for a* SEQ-MDP *that is a transformation of an* IDP-MDP *with $N > 1$, has an expected cost $V^{seq}$ which is bounded by $V^{seq} = V^* + \sum_{k=1}^{K}(\delta_k - \delta_1) + N(c_{N+1} - c_1)$, where $V^*$ is the optimal value for the* IDP-MDP.

*Proof.* We will make a constructive argument by providing a policy $\pi \in \Pi$ that realizes the bound $V^{seq}$. Consider a policy $\pi$ that starts with the lowest cost action, and offers the highest possible incentive. The policy proceeds by sequentially decreasing the offered incentive $\delta_k$ for alternate agent action $a_1$ until the agent rejects the offered incentive. Let this highest incentive at which the agent rejects be $\delta_{j-1}$. That means that the true incentive $t_1$ for alternate agent action $a_1$ is $\delta_j$. Since $a_1$ is the lowest cost and lowest reward alternate action, we know that $\delta_j$ will be accepted for all other alternate agent actions $i > 1$, since all such actions yield higher reward to the agent. Therefore we need not offer $\delta_j$ for any other agent actions. The policy then moves on to offering incentives for alternate agent action $a_2$, starting with offering $\delta_{j-1}$. The policy again continues to incrementally decrease the offered incentive until it reaches a reject, at which it then starts to provide incentives for $a_3$, and so on. The policy is similarly defined for all subsequent alternate agent actions. The defined policy $\pi$ operates only on the SEQ-MDP version of the IDP-MDP.

Note that after the principal makes an offer, either an incentive for a particular action, or a particular incentive for all actions can be eliminated. Since there can at most be $K$ accepts, and at most $N$ rejects (as at most one offer can be rejected for each action), there are at most $K + N$ steps.

Each of the at most $N$ rejected offers will cause the algorithm to incur at most $c_{N+1} - c_1$ additional cost compared to the optimal policy, since the principal will pay the cost of the default action $c_{N+1}$. Each offer that was accepted that is above the true incentive will result in an additional cost of at most $\delta_i - \delta_1$. Note that each incentive $\delta_i$ is offered only once if it is accepted for an action $a_n$. Therefore, the total cost is bounded by $V^{seq} = V^* + N(c_{N+1} - c_1) + \sum_{k=1}^{K}(\delta_k - \delta_1)$. This is true irrespective of the distribution over incentives. It can be easily seen that the policy $\pi \in \Pi$, as required. □

This bound does not use any information about the probability distribution. A tighter bound using the probability distribution can be obtained but is not presented here for ease of exposition. Also, an alternate algorithm with bound $V^* + K(c_{N+1} - c_1) + \sum_{i=1}^{N}(c_N - c_i)$ is possible by starting from action $a_N$ rather than $a_1$ and progressing to lower cost actions only after identifying the true incentive for the higher cost actions. An argument similar to the one made in Theorem 6.2 can be used to prove this bound. In next section, our empirical results show the optimal policy for a SEQ-MDP performs well in practice.

## 7 Experiments

We now empirically evaluate our algorithms.

In our simulations the cost of the default agent action was set to $c_{N+1} = 2$. For $N$ alternate agent actions, the cost to the principal of the agent taking action $a_n$ was set to $c_n = \left(\frac{n}{N}\right)^\eta$ where $\eta$ is a constant. For $K$ incentives, the value of the $k$-th incentive was $\delta_k = k/K$. Note the maximum cost of the alternate agent actions is 1, and the maximum incentive is 1. Therefore the cost to the principal of an agent accepting an incentive $c_n + \delta_k$ is always less than or equal than the cost to the principal of the agent's default action $c_{N+1}$, as our IDP model assumed (Section 4). The initial belief was set to a uniform joint distribution over the possible incentive-alternate agent action space: recall that this distribution must always respect the constraints that $t_i \geq t_j$ for $i < j$ since $R_{a_i} < R_{a_j}$. In each experimental run we fixed $K, N, \eta$ and the horizon $H$, and sampled a true hidden vector of incentives for the agent from the initial belief. We then executed the policy of each algorithm and recorded the total cost accumulated over $H$ steps. We simulated 1000 runs and averaged over each run's total sum of costs. We repeated this for 10 rounds (each of 1000 runs) and computed the standard deviation error bars for the average total cost.

We compare the performance of our algorithms to two natural methods. The first is a greedy algorithm. In the greedy algorithm, given a current state $B_S$ (with $S_n = (s_n, e_n)$), the principal offers the incentive $\delta_k$ for alternate action $a_n$ that minimizes the expected immediate cost to the principal,

$$[\delta_k, a_n] = \operatorname*{argmin}_{\delta_k, a_n} \sum_{k'=s_n}^{k} p^n_{k',S}(\delta_k + c_n) + \sum_{k'=k+1}^{e_n} p^n_{k',S}(c_{N+1}).$$

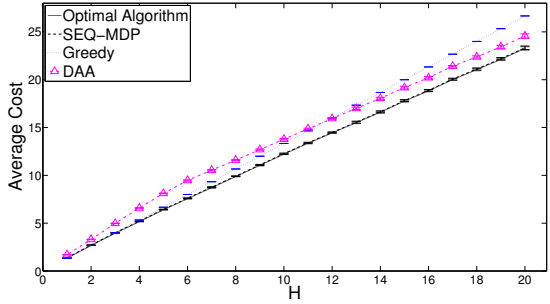

(a) Avg total cost over $H$ (w/1 std. error bars). $N=3, K=5$.

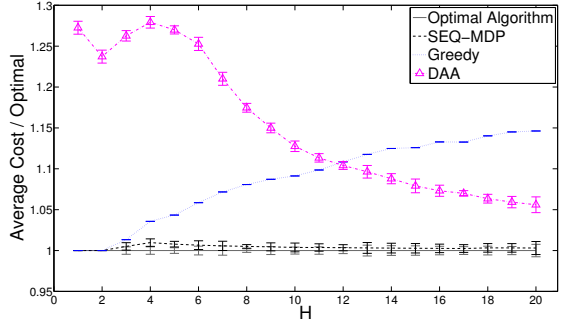

(b) Ratio of avg. total cost to optimal. $N=3, K=5$.

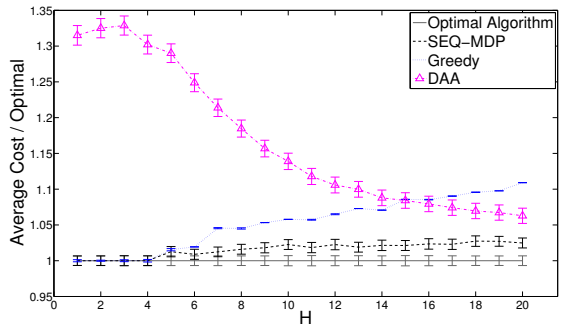

(c) Ratio of costs. $N=5, K=3$.

Figure 1: Average total cost for varying horizons $H$.

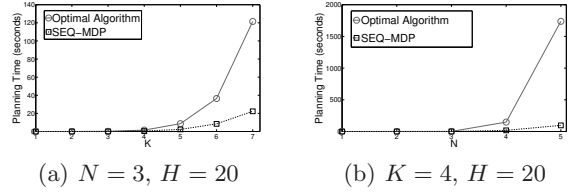

(a) $N=3, H=20$  (b) $K=4, H=20$

Figure 2: Planning time (sec) as a function of K or N for the optimal algorithm and SEQ-MDP.

The second method first uses a binary-search-like procedure, starting from $a_1$, to identify the true incentive for each alternate agent action. It then selects the alternate agent action, and incentive, with the lowest immediate cost, and then offers that for all remaining steps. We label this algorithm "diagnose-and-act (DAA)." It should be noted that DAA can be viewed as a first running a simple inverse reinforcement learning approach.

Figure 1(a) displays the average total cost incurred for different horizons for $N=3$ alternate agent actions, and $K=5$ incentives. Here we set $\eta=1$ and $N \leq 5$ but results from varying $\eta$ (from 0.75 to 1.25) and $N$ yielded similar results. As expected, the optimal policy performs best for all horizons. To better explore the difference between the algorithms, we replotted the results by dividing by the optimal algorithm's average total cost at each horizon $H$, resulting in a figure where the optimal algorithm is always 1, and all other algorithms are displayed as their ratio to the optimal performance (see Figure 1(b)). For low horizons the greedy algorithm performs very close to optimal, because there is little benefit to gathering information to find a lower cost alternate since there is little time to use such information. However, as $H$ increases, greedy performs significantly worse than the optimal algorithm. DAA algorithm performs poorly at low horizons because it tries to identify the optimal incentive for each action, and this can take longer than the available horizon to accomplish. At larger horizons DAA performs better. For very large horizons we expect that DAA would converge to the performance of the optimal policy: in such cases the loss incurred from not identifying the best action-incentive pair will likely become very high, meaning that it will be optimal to identify the minimal cost incentive.

The SEQ-MDP approach has very good performance. Indeed it is was visually and statistically indistinguishable from the optimal algorithm for all horizons $H$ for $K=5, N=3$. SEQ-MDP is not always identical to the optimal algorithm's performance: for example, in Figure 1(c) we display results for $K=3$ and $N=5$. Here there are a larger number of possible incentive-action combinations ($3^5$) relative to the available horizon, and so consecutively stepping through the alternate agent actions may not be the optimal strategy.

However SEQ-MDP performs very close to the optimal algorithm's performance, and it is significantly better than other approaches. The SEQ-MDP approximation also requires much less computation time than the optimal algorithm. Figures 2(a) and 2(b) compare the average computational time to execute a single run ($H=20$). Figure 2(a) fixes the number $N=3$ of alternate agent actions, and varies the number of offered incentives $K$, and Figure 2(b) fixes $K=4$ and varies $N$. The SEQ-MDP algorithm scales much better than the optimal algorithm in both cases.

## 8 Discussion and Extensions

So far we have focused on Incentive Decision Processes where a principal interacts with a myopic greedy agent whose preferences are hidden and static. Here we briefly consider two extensions of this model.

**Dynamic Preferences**: In many real applications the agent's preferences will change over time, and the principal may not know if the agent's preferences have changed. This significantly complicates the decision problem. One important common class of dynamic changes is where the preferences change locally. For example, a tenant's thermal preferences may locally shift during seasonal changes. Let us assume that if true incentive $t_n$ for alternate agent action $a_n$ is $\delta_k$, then $t_n$ remains the same with probability $\lambda_0$, shifts to $\delta_{k+d}$ with probability $\lambda_1$, and shifts to $\delta_{k-d}$ with probability $\lambda_2$ after each potential preference change[2]. We will restrict our attention to the case where the maximum number of times the agent might change its preferences is at most $L$. For simplicity we will also assume that we know the time epochs at which changes can occur, but our results also apply when we do not have prior information about these time epochs. Similarly, we will state our results for $N = 1$ but these results can be extended to $N > 1$.

Such dynamic preference IDPs can be represented as a POMDP with $O(K^3)$ states: in a static IDP there are $O(K^2)$ belief states; a change in preference shifts the non-zero elements of the belief state; there are at most $O(K)$ possible shifts; and as we don't know whether a shift occurs or not, this becomes the hidden state of our POMDP. Interestingly, we can reduce this POMDP to a finite state MDP:

**Theorem 8.1.** *A dynamic preference IDP (N=1) can be represented as a MDP with $O(K^{2(L+1)})$ states.*

This result is significant because we have mapped a non-deterministic POMDP to a finite state MDP. While a similar result is known to hold for deterministic POMDPs, it does not hold for general POMDPs, which typically map to a MDP with an infinite number of states. It is important to emphasize that this result holds for infinite horizon dynamic preference IDPs.

We believe that substantial further reductions in the state space may be possible with minimal performance loss since each possible change is local and small.

**Rational Strategic Agents:** Often the agent itself will seek to optimize its long term expected sum of rewards, reasoning about the potential future incentives the principal might provide if the agent chooses to accept or reject different offers. In such situations, the previous reduction to a MDP will no longer hold because the principal cannot trust that the agent's responses reflect its true hidden incentives.

Many of the policies we have considered so far will provide a higher incentive when an offer is rejected, and a lower incentive if the offer is accepted. Let us denote such policies $\pi^{simple}$. Such policies are strategy proof for a horizon of $H = 1$:

**Theorem 8.2.** *The agent will act truthfully for $H = 1$.*

Indeed it is fairly easy to see that acting truthfully will always maximize the agent's immediate reward. However, this does not hold for longer horizons.

**Theorem 8.3.** $\pi^{simple}$ *is not a strategy proof policy for arbitrary horizons.*

The proof provides an example of a $H = 2$ IDP where the agent can increase its reward by acting in discord with its true preferences. In the future we intend to develop algorithms for automatically constructing equilibrium decision policies for the principal and agent, and to design strategy proof policies.

**Other directions:** There are multiple other extensions to the Incentive Design Problems we have considered in this paper. Another interesting variant is when there are multiple agents (say $U$), and the principal can only provide incentives to 1 agent at each time step. It can be shown that an approach similar to binary search will lead to an $O((c_2 - c_1)U \log K)$ additive bound with respect to the optimal algorithm in case of single alternate action. Interestingly, this problem can also be reduced to multi-arm bandit (MAB) problem with superprocesses [8]. Though computing optimal policies for superprocess MABs is generally difficult, it may be possible to leverage the IDP structure to compute efficient algorithms. We also plan to use IDPs to construct personalized adaptive incentives for reducing energy consumption.

## 9 Conclusion

We have introduced Incentive Decision Processes, where the objective is to compute a decision policy for a principal to minimize its expected sum of costs by providing incentives to a myopic agent. If the agent's hidden preferences are static, then we can represent an IDP (either exactly for 1 alternate action, or approximately for multiple actions) as a polynomially-sized Markov decision process, instead of as a standard POMDP. Our empirical results showed our sequential decision theoretic techniques significantly outperform simpler comparison algorithms.

---

[2] $\lambda_0 + \lambda_1 + \lambda_2 = 1$.